\documentclass[pdflatex,sn-nature]{sn-jnl}

\makeatletter
\let\blx@rerun@biber\relax
\makeatother

\usepackage[british]{babel}
\usepackage[british]{datetime2}
\usepackage{graphicx}%
\usepackage{multirow}%
\usepackage{amsmath,amssymb,amsfonts}%
\usepackage{amsthm}%
\usepackage{mathrsfs}%
\usepackage[title]{appendix}%
\usepackage{xcolor}%
\usepackage{textcomp}%
\usepackage{manyfoot}%
\usepackage{booktabs}%
\usepackage{algorithm}%
\usepackage{algorithmicx}%
\usepackage{algpseudocode}%
\usepackage{listings}%
\usepackage{siunitx}%
\usepackage{xfrac}%
\usepackage[style=nature,defernumbers=false]{biblatex}
\usepackage{lineno}
\usepackage{csquotes}

\renewcommand{\d}{\mathop{}\!{\mathrm{d}}}

\newcommand{\iu}{{i\mkern1mu}}
\newcommand{\mycopyright}{\textsuperscript{\tiny{®}}}

\begin{document}

\title[Article Title]{Acousto-Optic Modulation in Ambient Air}

\author*[1]{\fnm{Yannick} \sur{Schrödel}}\email{yannick.schroedel@desy.de}
\author[2]{\fnm{Claas} \sur{Hartmann}}
\author[1]{\fnm{Jiaan} \sur{Zheng}}
\author[1]{\fnm{Tino} \sur{Lang}}
\author[3]{\fnm{Max} \sur{Steudel}}
\author[2]{\fnm{Matthias} \sur{Rutsch}}
\author[1,4,5]{\fnm{Sarper H.} \sur{Salman}}
\author[6]{\fnm{Martin} \sur{Kellert}}
\author[6]{\fnm{Mikhail} \sur{Pergament}}
\author[7]{\fnm{Thomas} \sur{Hahn-Jose}}
\author[2]{\fnm{Sven} \sur{Suppelt}}
\author[2]{\fnm{Jan Helge} \sur{Dörsam}}
\author[3]{\fnm{Anne} \sur{Harth}}
\author[1,8]{\fnm{Wim P.} \sur{Leemans}}
\author[6,8,9]{\fnm{Franz X.} \sur{Kärtner}}
\author[1]{\fnm{Ingmar} \sur{Hartl}}
\author[2]{\fnm{Mario} \sur{Kupnik}}
\author[1,4,5]{\fnm{Christoph M.} \sur{Heyl}}

\affil*[1]{\orgdiv{FS-LA}, \orgname{Deutsches Elektronen-Synchrotron DESY}, \orgaddress{\city{Hamburg}, \country{Germany}}}

\affil[2]{\orgdiv{Measurement and Sensor Technology}, \orgname{Technische Universität Darmstadt}, \orgaddress{\city{Darmstadt}, \country{Germany}}}

\affil[3]{\orgdiv{Department of Optics and Mechatronics}, \orgname{Hochschule Aalen}, \orgaddress{\city{Aalen}, \country{Germany}}}

\affil[4]{\orgname{Helmholtz-Institute Jena}, \orgaddress{\city{Jena}, \country{Germany}}}

\affil[5]{\orgname{GSI Helmholtzzentrum für Schwerionenforschung GmbH}, \orgaddress{\city{Darmstadt}, \country{Germany}}}

\affil[6]{\orgdiv{Center for Free-Electron Laser Science CFEL}, \orgname{Deutsches Elektronen-Synchrotron DESY}, \orgaddress{\city{Hamburg}, \country{Germany}}}

\affil[7]{\orgname{inoson GmbH}, \orgaddress{\city{St. Ingbert}, \country{Germany}}}

\affil[8]{\orgdiv{Department of Physics}, \orgname{University of Hamburg}, \orgaddress{\city{Hamburg}, \country{Germany}}}

\affil[9]{\orgname{The Hamburg Centre for Ultrafast Imaging}, \orgaddress{\city{Hamburg}, \country{Germany}}}


\abstract{ 
    Control over intensity, shape, direction, and phase of coherent light is essential in numerous fields, reaching from gravitational wave astronomy over quantum metrology and ultrafast sciences to semi-conductor fabrication. 
    Modern laser optics, however, frequently demands parameter regimes where either the wavelength or the optical power restricts control due to linear absorption, light-induced damage or optical nonlinearity. 
    The properties of solid media, upon which most photonic control schemes rely, impose these limitations. 
    We propose to circumvent these constraints using gaseous media tailored by high-intensity ultrasound waves. 
    We demonstrate a first implementation of this approach by deflecting ultrashort laser pulses using ultrasound waves in ambient air, entirely omitting transmissive solid media. 
    At optical peak powers of $\SI{20}{\giga\watt}$ exceeding previous limits of solid-based acousto-optic modulation by about three orders of magnitude, we reach a deflection efficiency greater than $50\%$ while preserving excellent beam quality. 
    Our approach is not limited to laser pulse deflection via acousto-optic modulation: 
    gas-phase photonic schemes controlled by sonic waves can prospectively be translated to various optical methods, e.g., lenses or waveguides, rendering them effectively invulnerable against damage and opening up new spectral regions.
}



\maketitle

\newpage
\begin{refsegment}

\textbf{
Control over intensity, shape, direction, and phase of coherent light \supercite{yu_light_2011,drescher_extreme-ultraviolet_2018,tirole_double-slit_2023,balla_ultrafast_2023,berti_extreme_2022} is a cornerstone of numerous fields, reaching from gravitational wave astronomy \supercite{abbott_observation_2016} over quantum metrology \supercite{schnabel2010quantum,nichol2022elementary} and ultrafast sciences \supercite{lindroth_challenges_2019,biegert_attosecond_2021} to semiconductor fabrication \supercite{tallents2010lithography}.
Modern photonics, however, frequently demands parameter regimes where either the wavelength or the optical power restricts control due to e.g., absorption or damage. 
The properties of solid media, upon which most photonic control schemes rely, impose limitations. 
Here we circumvent these limitations, employing gas media tailored by intense ultrasound waves. 
We demonstrate a first implementation of this approach by angularly deflecting ultrashort, high power laser pulses using ultrasound waves in ambient air, entirely omitting transmissive solid media. 
At optical peak powers of \boldmath{$\SI{20}{\giga\watt}$} exceeding the limits of solid-state based acousto-optic modulation by about three orders of magnitude \supercite{riesbeck_aom_2009,mazur_acousto-optic_2020}, we reach a pulse deflection efficiency greater than 50\% while preserving excellent beam quality.
Our approach is not limited to laser pulse deflection via acousto-optic modulation. 
Gas-phase photonic schemes controlled by sonic waves can prospectively be translated to various optical methods, rendering them effectively invulnerable against damage and expanding into new spectral regions.
}

\vspace{5mm}

The main macroscopic mechanism governing the propagation of light in a medium is the medium's refractive index, $n$. 
When an optical wave propagates through a medium, its propagation speed changes from $c_0$, the speed of light in vacuum, to $c_0 / n$. 
This ultimately leads to a change in phase and intensity. 
For example, an interface between different refractive indices causes a change in propagation direction of light both in reflection and transmission \supercite{yu_light_2011}. 
The strength of this effect depends on the difference between the refractive indices $\Delta n$ and the incident angle\supercite{fresnel_augustin_note_1821}. 
Large refractive index differences on the order of $\Delta n \sim 0.5$ can be reached at the boundary between gases and transparent solid for most optical wavelengths \supercite{ciddor_refractive_1996}. 
This is a key reason why bulk media are used almost exclusively for optical elements such as lenses, mirrors, waveguides and many more. 

However, with the rapid progress in high peak-power laser technology\supercite{mourou_nobel_2019} and applications\supercite{albert_2020_2021,biegert_attosecond_2021,kodama_fast_2002} reaching into novel wavelength regimes\supercite{lindroth_challenges_2019,maroju_attosecond_2020}, established solid-based control schemes face severe limitations. 
Compared to gases, glasses are transmissive only in a relatively small spectral range, they restrict optical peak and average power through light-induced damage \supercite{natoli_laser-induced_2002} as well as thermal lensing \supercite{gordon_longtransient_1965} and cause losses at boundary layers. 
For intense or ultrashort pulses, additional restrictions arise due to dispersion and nonlinear optical effects such as self-focusing \supercite{kelley_self-focusing_1965}. 
One powerful route to circumvent some of these limitations has been opened by meta optics \supercite{kamali_review_2018,ossiander_extreme_2023}, relying on nanostructured dielectric media. 

An entirely different route is the employment gaseous photonic media.
In contrast to solids, gases are immune to damage and support about three orders of magnitude higher peak powers at very little dispersion within large spectral regions. 
Their refractive index, however, is very close to \num{1}, limiting $\Delta n$ for gas-based photonic systems. 
In addition, creating a static refractive index boundary in gases poses a technical challenge. 
However, in the limit of small incident angle (grazing incidence), light reflection can still occur even for small $\Delta n$. 
In nature, this phenomenon is well-known: in a mirage, layers of air at different temperature levels, while $\Delta n$ is in the order of only $10^{-5}$, can alter the optical path substantially \supercite{ciddor_refractive_1996}. 
Applying similar principles, gas-phase refractive elements \supercite{drescher_extreme-ultraviolet_2018} such as lenses \supercite{grulkowski_acousto-optic_2009} as well as gratings using multiple plasma layers \supercite{michine_ultra_2020} have been developed. 

Strikingly, more advanced gas-based schemes enabling superior control options including acousto-optic modulation (AOM) have not entered the photonics field until today. 
Here we demonstrate a novel light control method by employing intense ultrasound waves in air. 
To achieve this, we engineer a transmissive optical Bragg-grating by periodic sinusoidal pressure modulation \supercite{phariseau_diffraction_1956}, enabling efficient AOM in ambient air. 

\begin{figure}
    \begin{center}
        \includegraphics{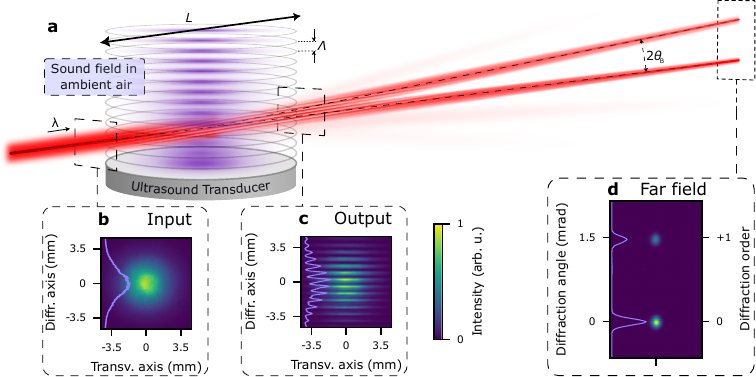}
        \caption{\label{fig1} \textbf{Schematic of ultrasound-assisted angular deflection of a laser beam in air.} 
            \textbf{a}, A laser beam (red) in a gaseous medium entering a sound field at a shallow Bragg angle relative to the sound field wave fronts (purple). 
            Acousto-optic modulation leads to efficient deflection of the incident beam at twice the Bragg angle and at higher diffraction orders carrying minimal energy. 
            \textbf{b}-\textbf{d}, Camera-measured beam profiles and corresponding centred line-outs along the diffraction axis of an approximately Gaussian-shaped incident beam (\textbf{b}), directly after the acousto-optic interaction, exhibiting interference-induced fringes (\textbf{c}) and in the far field (\textbf{d}). 
            The  data were recorded in a multi-pass setup enabling an increased interaction length.
            }
    \end{center}
\end{figure}

\subsection*{Acousto-optics: From solids to gases}\label{sec:aosolidtogas}

The field of acousto-optics (AO) describes the interaction between optic and acoustic waves. 
Upon interaction of the two waves, quantized as a photon and a phonon, they can scatter \supercite{klein_unified_1967}. 
Momentum and energy conservation lead to partial redirection and a frequency shift. 
This enables direct control over the optical frequency, propagation direction, phase, and intensity through the acousto-optic effect via the readily accessible acoustic frequency $F$ (\SI{100}{\mega\hertz} range for solid media) and the acoustic power.
The acoustic wavelength $\Lambda$ appears as a periodic modulation of the optical refractive index with a magnitude in the order of $\Lambda = V/F \approx \SI{10}{\micro\meter}$, where $V$ is the speed of sound. 
When the refractive index modulation depth $\Delta n \left(x, t\right)$ is sufficiently high and the interaction length $L$ is large enough \supercite{goutzoulis_design_1994}, optimal conversion into the first diffraction order $m=+1$ can be achieved by phase-matching of the diffracted and the incident ($m=0$) waves. 
This interaction type is called the Bragg regime, where scattering into higher diffraction orders is minimized. 
Phase-matched conversion is reached when the angle $\theta$ between incident beam propagation direction and the acoustic wave-fronts fulfils the Bragg condition 
\begin{equation}\label{eq:bragg_condition}
    \sin \theta_\mathrm{B} = \frac{\lambda F}{2 V}.
\end{equation}

Diffraction efficiencies into the ($m = +1$)-order close to 100\% have been achieved in solid-state AO modulators, and these are widespread in modern photonics. 
Applications include active Q-switching and cavity dumping by loss modulation \supercite{plaessmann_subnanosecond_1993}, stabilization of carrier-envelope phase \supercite{jones_carrier-envelope_2000} and optical gating \supercite{schwenger_high-speed_2012}. 
Employed as AO tunable filter \supercite{harris_acousto-optic_1969} and as programmable dispersive filters \supercite{tournois_acousto-optic_1997,shim_femtosecond_2006}, AO devices have been used to spectrally select and shape the phase of ultrashort pulses. 
The AO medium used to carry the acoustic wave determines the characteristics of the AOM. 
A number of crystalline \supercite{yano_tellurite_1971,abrams_acoustooptic_1970,pinnow_lead_1969,krause_low_1970} and liquid \supercite{nagai_acousto-optical_1977,nikitin_acousto-optic_2017,szulzycki_generation_2016} materials have been used, with TeO$_2$ and crystalline Germanium being the most common ones. 

Employment of a gas as AO medium has so far only been reported for the far infrared using a high-pressure gas vessel \supercite{durr_acousto-optic_1986} and to modulate intra-cavity losses for He-Ne and Argon-ion lasers in cylindrical configurations \supercite{grulkowski_acousto-optic_2009}. 
This can be attributed to the acoustic properties of gases setting severe technical challenges: 
The first one is the low $\Delta n$ in gases, arising from the pressure modulation $\Delta p$ and the emerging temperature differences with $\Delta n \sim \Delta p \times \left(n-1\right)$. 
Even for high sound pressure levels ($\mathrm{SPL}$) of e.g. \SI{140}{\deci\bel}, $\Delta n$ in air reaches values in the order of only $10^{-7}$. The second challenge is the high ultrasound field absorbance for even comparatively low $F$, effectively limiting the acoustic frequencies to values below a few \si{\mega\hertz}\supercite{bond_absorption_1992}. 
In order to ensure efficient AOM in gases, two requirements must be met: First, to ensure strong scattering \supercite{klein_unified_1967}, $\Delta n$ must be large enough, and a shallow incidence angle must be used to take advantage of high Fresnel reflectance at grazing incidence. 
Separability of the diffracted and transmitted beams set a lower bound on the incidence angle and thus on $F$. 
For example, a Gaussian beam with a $\sfrac{1}{\mathrm{e}^2}$-radius of $w_0=\SI{3.5}{\milli\meter}$ in the NIR ($\lambda = \SI{1}{\micro\meter}$) has a diffraction-limited beam divergence of $\theta_0 \approx \SI{100}{\micro\radian}$. 
Using Eq. \ref{eq:bragg_condition}, we determine that a beam deflection angle of $2 \theta_\mathrm{B} \geq 4 \theta_0$ requires $F \geq \SI{135}{\kilo\hertz}$ for AOM in air. 
Second, the transducer length $L$ must be optimized with respect to $\Delta n$ and the angular acceptance\supercite{goutzoulis_design_1994,eddie_h_young_design_1981,maydan_acoustooptical_1970}. 
While a small $\Delta n$ demands a large AO interaction length, the angular acceptance shrinks with growing $L$. 
Fig. \ref{fig1}\textbf{a} illustrates the principle of AOM employing a shallow angle in air, where spatial separation of transmitted and deflected beams is enabled in the far field. 

\subsection*{Basic characteristics of gas-phase acousto-optic modulation}\label{sec:basicgasaom}

\begin{figure}
    \begin{center}
        \includegraphics{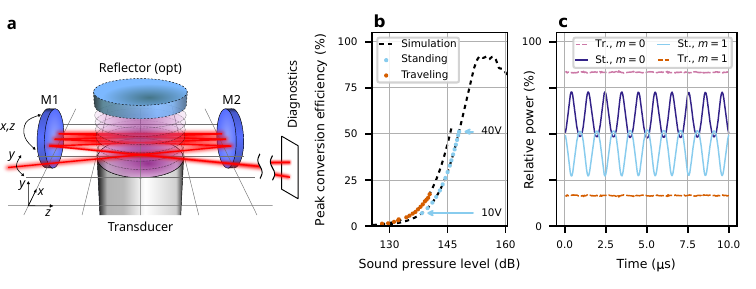}
        \caption{\label{fig2}\textbf{Experimental setup and dependence of deflection efficiency on acoustic power and time.}
        \textbf{a}, Schematic of the AO-interaction geometry. 
        Mirrors M1 and M2 of the multi-pass folding geometry are indicated in blue. 
        The resulting beam path is marked red. 
        The optional acoustic reflector is indicated in light blue. 
        \textbf{b}, Maximum achieved conversion efficiency into the first diffraction order ($m = +1$) as a function of SPL for both standing acoustic wave (blue dots) and travelling acoustic wave without the optional reflector (orange dots) displayed together with simulated signals (black dashes). 
        Standing acoustic wave data points are fitted to simulated data. 
        The applied driver supply voltage boundaries are indicated. 
        \textbf{c}, Time evolution of relative powers of diffracted ($m = +1$) and transmitted ($m = 0$) orders of both standing (St., light/dark blue lines) and travelling (Tr., orange/purple dashes) wave configurations.}
    \end{center}
\end{figure}

Our experimental demonstrations of gas-based AO are carried out in ambient air using ultrashort laser pulses centred at \SI{1030}{\nano\meter}. 
Fig. \ref{fig2}\textbf{a} shows a schematic of the experimental setup employing all-reflective bulk optical elements. 
An incident laser beam passes seven times across an ultrasound field in order to increase the interaction length while keeping the ultrasound transducer at a reasonable size. 
An ultrasound transducer generates the ultrasound field and an optional planar reflector mounted to reflect the acoustic waves enhances the $\mathrm{SPL}$ by creating a standing ultrasound wave.
After AO interaction, the laser beams are routed to diagnostics enabling characterization of beam quality and power in all diffraction orders. 

The light source for our first experimental demonstration is a fibre-coupled mode-locked laser delivering ultrashort pulses ($\approx \SI{150}{\femto\second}$) at a pulse repetition rate of \SI{54}{\mega\hertz}. 
For this experiment, we use an average output power of a few tens of \si{\milli\watt}, preventing any nonlinear effects. 
Beam delivery via a polarization-maintaining single-mode fibre ensures an excellent spatial beam quality as shown in Fig. \ref{fig1}\textbf{b}, corresponding to beam quality factor $M^2 = 1.04$ in both axes. 
The collimated beam is subsequently sent into the ultrasound field at $\theta_\mathrm{B} \approx \SI{0.75}{\milli\radian}$ relative to the sound wave fronts. 
Following the AO interaction, the output beam exhibits interference fringes caused by the interference of all transmitted diffraction orders (Fig. \ref{fig1}\textbf{c}). 
In the far field (Fig. \ref{fig1}\textbf{d}), the diffraction orders are separated by $2 \theta_\mathrm{B} \approx \SI{1.5}{\milli\radian}$, an angle almost an order of magnitude larger than the beam divergence. 
Transmitted and diffracted beams display an $M^2$ of $1.15$ in both axes for both beams ($m = 0$, $m = +1$), demonstrating that an excellent beam quality is maintained upon AO interaction. 
Figure \ref{fig2}\textbf{b} displays the measured diffraction efficiency into the first order as a function of SPL for two configurations: with (red) and without (blue) acoustic reflector. 
In addition, the numerically calculated efficiency is displayed for both configurations. The simulations employ finite element methods for two-dimensional acoustics simulation and nonlinear split-step-Fourier methods for $2+1$ dimensional optical beam propagation (see Methods \ref{methods:numerical}). 
The sound field simulations are based on transducer surface velocity data extracted via laser-Doppler-vibrometry.
We adjust the ultrasound field strength by scanning the supply voltage of the driver of the ultrasound transducer. 
The voltage is tied to the transducer's surface velocity and therefore to the induced $\mathrm{SPL}$. 
The $\mathrm{SPL}$ is difficult to determine as acoustic microphones typically impact the sound field, particularly for a standing wave configuration. 
We therefore choose to calibrate the $\mathrm{SPL}$ for the standing wave case by fitting the measured data points to the simulated efficiency curves. 
For the travelling wave configuration, we use a commercial optical microphone to calibrate the $\mathrm{SPL}$. 
The efficiency trace displayed in Fig \ref{fig2}\textbf{b} (orange dots) show excellent agreement with the numerical simulations (dashes). 
The simulations indicate that the travelling wave configuration exhibits a slightly higher diffraction efficiency within the explored $\mathrm{SPL}$ range. 
This effect may be explained by a more homogeneous sound field for the travelling wave case and/or inaccuracies in the numerical model for the resonant acoustic wave. 
The maximum efficiency is obtained using the acoustic reflector yielding a diffraction efficiency into the first order exceeding 50\%. 
In comparison, the travelling wave without reflector can only deflect up to 20\% of the incident power. 
Strikingly, the numerical simulation indicates much higher diffraction efficiencies with further increased sound pressure levels. 

The two configurations, employing both travelling and standing waves, exhibit a vastly different temporal behaviour. 
The temporal characteristics were measured using photo diodes and are shown in Fig. \ref{fig2}\textbf{c}, corresponding to the data points obtained at maximum voltage displayed in Fig. \ref{fig2}\textbf{b}. 
The travelling wave case displays a constant relative optical power for both diffraction orders. 
On the contrary, the configuration using an acoustic resonator modulates the laser beams, reflecting the temporal amplitude oscillation of a standing wave with an oscillation frequency equal to $2 F \approx \SI{1}{\mega\hertz}$. 
Although a perfect standing wave expectedly exhibits periodic instants in time were the ultrasound field vanishes completely, we observe that the diffracted beam does not completely vanish. 
We attribute this to the acoustic field not exhibiting exclusively a fundamental transverse mode but showing phase variations across the radial extent of the ultrasound resonator. 

While resonant AOM enables modulation and switching of optical signals at $2F$, non-resonant (travelling wave) AOM offers more versatile temporal modulation options at reduced rates.
Both the speed of sound $V$ and the beam diameter, as well as the transducer's response define the AO rise time, which determines these rates. 
In our experiment, we determine the 10\% to 90\% rise time for non-resonant AOM to be \SI{32}{\micro\second} (see Methods \ref{methods:temporal}), corresponding to a possible modulation frequency in the \SI{100}{\kilo\hertz} range.
This may be improved by reducing the transducer rise time and/or employing gases with higher $V$.

\subsection*{Gigawatt-scale acousto-optic modulation}\label{sec:gigawatt}

\begin{figure}
    \begin{center}
        \includegraphics{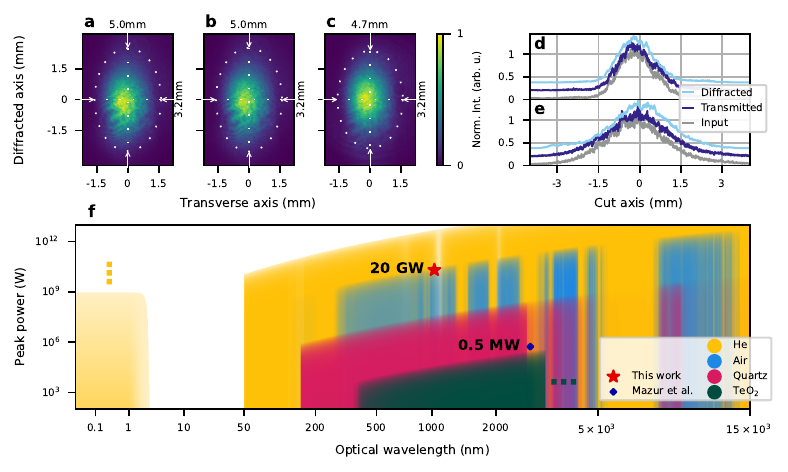}
        \caption{\label{fig3}\textbf{Acousto-optic diffraction of ultra-short laser pulses at high peak power.} 
        \textbf{a}-\textbf{c}, Measured spatial profiles of the input (\textbf{a}), transmitted (\textbf{b}) and diffracted (\textbf{c}) beams. 
        The major and minor axes, the $\sfrac{1}{\mathrm{e}^2}$-diameters and the circumferences of fitted elliptical Gaussian beams are annotated (white dashes and arrows). 
        \textbf{d}, \textbf{e}, Centred line-outs along transverse (\textbf{d}) and diffracted (\textbf{e}) axes are shown, normalized and vertically offset to improve readability. 
        \textbf{f}, Possible parameter regimes of gas (yellow: He, blue: air) and bulk-based (purple: quartz/ fused silica, green: TeO$_2$) AO media, limited by critical power and linear absorption, displayed together with the here reported result (red star). 
        Areas where data was not available are indicated using dotted line marks.
        Blue diamond: Ref. \supercite{mazur_acousto-optic_2020} using a KG(WO$_4$)$_2$ crystal as AO medium.
        For data sources and details see Methods \ref{methods:parameter}.}
    \end{center}
\end{figure}

In a second experiment using the same optical setup, we employ ultrashort (\SI{760}{\femto\second}) pulses with a pulse energy of up to \SI{15.2}{\milli\joule} and a corresponding peak power of up to \SI{20}{\giga\watt} delivered by a high-power laser amplifier \supercite{lang_high_2022}. 
The laser operates in burst mode (see Methods \ref{methods:temporal}) with an intra-burst average power of \SI{3.5}{\kilo\watt}.
Fig. \ref{fig3}\textbf{a} displays a camera image of the transmitted beam where no AO-interaction occurred (ultrasound transducer off), i.e., the laser beam simply propagates through the setup using the guiding mirrors. 
Fig. \ref{fig3}\textbf{b} shows the transmitted beam profile with active transducer. 
Finally, Fig \ref{fig3}\textbf{c} shows the spatially separated diffracted order at a diffraction efficiency of about 25\%. 
The ultrasound transducer was operated at a lower $\mathrm{SPL}$ compared to the maximum values shown in Fig. \ref{fig2}\textbf{b} limited by the driving electronics 
hardware available at the time of the experiment. 
The measured efficiencies matched those obtained using the fibre-coupled light source. 
In this experiment, the diffracted and transmitted order are spatially separated by \SI{16.5}{\meter} beam propagation of a low power fraction of the output beam behind the AO-interaction and measured via a lens telescope and a camera. 
Fig. \ref{fig3}\textbf{d}, \textbf{e} show centred line-outs of the beam profiles along the transverse (Fig. \ref{fig3}\textbf{d}) and diffraction (Fig. \ref{fig3}\textbf{e}) axes, respectively. 
The very similar beam shapes and sizes indicate preservation of the beam quality. 

The peak power is limited by the onset of self-focusing expected when exceeding the critical power of air (about \SIrange{5}{10}{\giga\watt})\supercite{liu_direct_2005}.
We observed this onset in our experiments at a maximum peak power of \SI{20}{\giga\watt} by a slightly reduced beam diameter following about four meters of beam propagation. 
Fig. \ref{fig3}\textbf{f} indicates the parameters peak power and optical wavelength used in our experiment (red star) together with estimated parameter regimes supporting AOM in helium (yellow shading) and air (blue shading). 
The commonly used bulk AOM materials quartz/ fused silica (purple shading) and TeO$_2$ (green shading) are shown for comparison along with reported values of a record peak power obtained using bulk AOM \supercite{mazur_acousto-optic_2020} employing KG(WO$_4$)$_2$ (crystals blue diamond). 
Critical power and linear absorption (see Methods \ref{methods:parameter}) define the displayed limits. 
Possible limitations arising due to geometrical restrictions such as a minimum diffraction angle supporting beam separation are not taken into account. 
The figure clearly shows the superior peak power performance of the demonstrated gas-AOM, exceeding earlier records by more than three orders of magnitude \supercite{riesbeck_aom_2009}. 
In addition, the figure indicates the potential to extend gas-based photonic devices such as AOM technology into the ultraviolet and beyond, as well as into the mid-infrared. 

\subsection*{Conclusion and outlook}\label{sec:conclusion}

In this work, we have demonstrated efficient and beam quality-preserving deflection of \si{\giga\watt}-scale laser pulses via AOM in ambient air while entirely omitting transmissive bulk optics. 
Our results demonstrate a major peak power-boost for AOM, limited by the critical power of the gas-phase AO medium. 
Considering lighter gases and/or gas-media at reduced pressure, gas-phase AOM can prospectively be scaled further, approaching the Terawatt regime. 
Contrarily, the usage of heavier gases with larger refractive indices and/or higher gas densities may enable diffraction efficiencies exceeding 50\% dramatically. 
Alternatively, more advanced ultrasound transducer schemes targeting increased $\mathrm{SPL}$ could enable an efficiency boost as indicated by our numerical simulations. 
In addition, using gases as AO media opens a route towards AOM in new spectral regions: 
Ambient air is transmissive in ranges from about \SI{250}{\nano\meter} towards the near and mid-infrared, thus exceeding the transmission band of solids dramatically. Noble gases such as He can prospectively support an even wider spectral range, including the ultraviolet region down to about \SI{50}{\nano\meter} and the transparent window for photon energies above \SI{1}{\kilo\electronvolt}. 

Gas-phase AOM can thus be expected to open entirely new parameter regimes for AOM technology with great prospects for applications including optical switches, beam splitters, phase modulators, spectral filters and many more. 
Moreover, the concept of employing gas-phase media tailored via intense ultrasound waves can be further expanded beyond AOM, opening the door to versatile gas-based photonics. 
These may include lenses, wave guides, optical gratings and considerably more, translating photonic devices from solid-state media to the gas-phase, thereby opening a new field: gas-phase Sono-Photonics. 

\printbibliography[title={References},segment=1,heading=subbibliography]
\end{refsegment}

\backmatter

\subsection*{Supplementary Information}

\bmhead{Acknowledgments}

We acknowledge Deutsches Elektronen-Synchrotron DESY (Hamburg, Germany) and Helmholtz-Institute Jena (Jena, Germany), members of the Helmholtz Association HGF for support and/or the provision of experimental facilities.

\subsection*{Declarations}

\bmhead{Funding}
\begin{itemize}
    \item DESY Generator Program, Deutsches Elektronen-Synchrotron DESY
    \item Klaus-Tschira Boost Fund (GSO/KT 29), Klaus Tschira Foundation
    \item CZS Wildcard (P2022-03-018), Carl Zeiss Foundation
    \item Seventh Framework Program (FP7) FP7/2007-2013 European Research Council (ERC) Synergy Grant (609920)
\end{itemize}

\bmhead{Competing Interest Declaration}
THJ is CEO of inoson GmbH, Sophie-Krämer-Straße 8, 66386 St. Ingbert.

CMH and TL have co-filed a (pending) patent on \DTMdate{2020-10-27} entitled acousto-optical modulator apparatus and method of acousto-optically deflecting a laser beam published as EP3992707A1 and US2022128883A1. 
Applicants for this patent are Deutsches Elektronen-Synchrotron DESY and GSI Helmholtzzentrum für Schwerionenforschung GmbH, inventors are Christoph Heyl and Tino Lang.
This patent involves an acousto-optical modulator apparatus for acousto-optically deflecting a laser beam and comprises a gas-filled volume including a working gas.

The other authors declare that they have no competing interests.

\bmhead{Authors' Contributions}
\begin{itemize}
    \item Conceptualization: YS, CH, TL, THJ, MK, CMH
    \item Methodology: YS, CH, TL, MR, THJ, SS, AH, IH, MK, CMH
    \item Software: YS, CH, TL, MR 
    \item Validation: YS, CH, TL, MR, JHD, CMH
    \item Formal Analysis: YS, CH, TL, MR
    \item Investigation: YS, CH, TL, JZ, MS, MR, SHS, SS, MK, JHD
    \item Resources: TL, JZ, MP, THJ, FXK, IH, MK, CMH 
    \item Data Curation: YS, CH
    \item Writing — original draft: YS, CMH 
    \item Writing — review and editing: YS, CMH and all other co-authors
    \item Visualization: YS
    \item Supervision: AH, WPL, IH, MK, CMH 
    \item Project administration: CMH
    \item Funding acquisition: AH, MK, CMH
\end{itemize}

\begin{appendices}

\section{Methods}\label{sec:methods}
\begin{refsegment}

\subsection{Numerical Methods}\label{methods:numerical}

\begin{figure}[H]
    \begin{center}
        \includegraphics{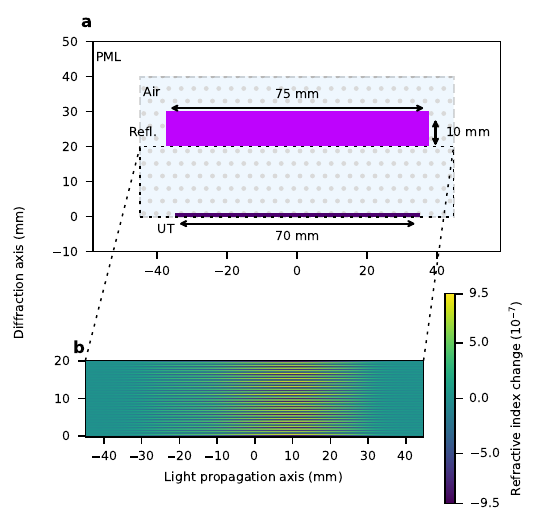}
        \caption{\label{fig:ed1}\textbf{Schematic of the geometry used for the acoustic simulation.} 
        \textbf{a}, A perfectly matched layer (PML) surrounds the air-filled region. 
        UT is the ultrasound transducer with an active area which exhibits a normal surface velocity distribution characterized using LDV measurements (see Fig. S2). 
        The optional reflector is indicated (Refl., purple area). 
        \textbf{b}, The resulting refractive index modulation obtained from the COMSOL model and the LDV measurements.}
    \end{center}
\end{figure}

We performed in-depth numerical simulations of the underlying acoustic and optic phenomena resembling the presented experimental setup. 
The simulation consists of two building blocks: 
First, the sound field is simulated using COMSOL Multiphysics\mycopyright. 
Then, this sound field is converted to a refractive index field. 
This refractive index field is used as an input to the optical simulation which is performed using the laser pulse propagation and nonlinear optics software Chi3D \supercite{lang_impact_2013}. 
The acoustic model uses finite element methods (FEM) \supercite{turner_stiffness_1956} of COMSOL Multiphysics\mycopyright. 
The large ratio between the acoustic wavelength of \SI{0.7}{\milli\meter} and the diameter of the reflector of \SI{75}{\milli\meter} requires a large grid. 
For this reason, a two-dimensional model is employed. 
The geometry considered for this model is sketched in Extended Data Fig. \ref{fig:ed1}. 
It takes into account an ultrasound transducer (UT) implemented as an acoustic aperture with a diameter of \SI{70}{\milli\meter} and, optionally, a reflector with a thickness of \SI{10}{\milli\meter} and a diameter of \SI{75}{\milli\meter}. 
The distance between the reflector and the UT is \SI{19}{\milli\meter}. 
The sound field is generated by considering a periodic displacement of the acoustic aperture represented by a spatially non-uniform distribution of the normal velocity of the UT surface. 
In order to closely resemble our experiment, measurements of the UT surface velocity in resonance are conducted using a Laser-Doppler vibrometre (LDV) \supercite{rothberg_international_2017}. 

\begin{figure}[H]
    \begin{center}
        \includegraphics{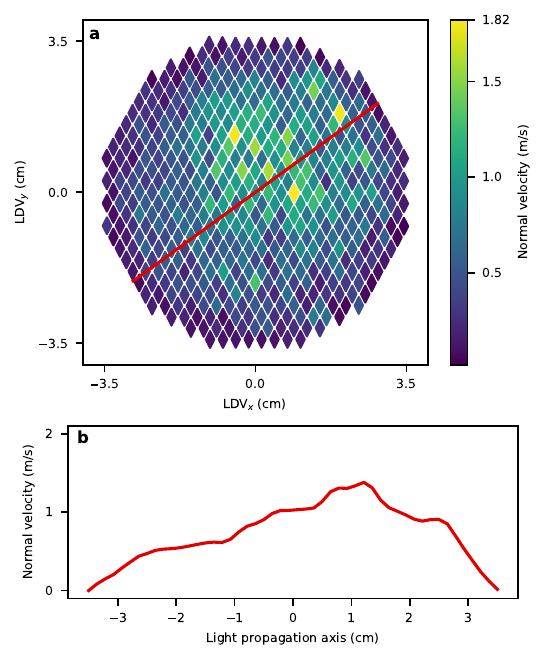}
        \caption{\label{fig:ed2}\textbf{Raw data of the LDV measurements of the employed ultrasound transducer.}
        \textbf{a}, Spatial distribution of the normal surface velocity of the ultrasound transducer. 
        The red line indicates the approximate line-out location used for the 2D acoustic simulation. 
        \textbf{b}, The extracted one-dimensional surface velocity along the red line displayed in \textbf{a}.}
    \end{center}
\end{figure}

Extended Data Fig. \ref{fig:ed2}\textbf{a} shows the two-dimensional results of the LDV measurements without reflector. 
Extended Data Fig. \ref{fig:ed2}\textbf{b} shows the result of a line scan along the two-dimensional transducer aperture such that it approximates the position of the modulated optical beams in our experiments inside the transducer-reflector arrangement. 
This one-dimensional data set is then used as input to model the sound field. 
Resembling the experimental conditions, air at \SI{1000}{\hecto\pascal} pressure, \SI{21.1}{\celsius} and 42.5\% relative humidity is used as material and a perfectly matched layer (PML) is placed to complete the model. 
This ensures absorption of all emitted waves leaving the transducer-reflector arrangement, representing an infinite propagation space for the acoustic waves. 
The ultrasound wave propagation simulations are based on solving the Helmholtz equation and include linear acoustic propagation and thermoviscous effects. 
We then convert the arising pressure field to a refractive index modulation field $\Delta n$.
Here, we consider influences from variation in pressure from normal pressure $\left(\Delta p / p_0\right)$ and temperature from ambient temperature $\left(\Delta T / T_0\right)$ using the modified Edlén equation \supercite{edlen_refractive_1966,birch_correction_1994}.
We linearly approximate the modified Edlén equation around the ambient conditions with refractive index $n$ as 
\begin{equation}
    \Delta n = \left(n-1\right) \times \left[ \Delta p / p_0 + \Delta T / T_0 \right].
\end{equation}
Extended Data Fig. \ref{fig:ed1} shows the emerging $\Delta n$-field used in our simulations and indicates the position of the UT and reflector. 

The optical simulations employ a split-step Fourier algorithm incorporating linear and nonlinear laser pulse propagation effects. 
This includes dispersion, self-phase-modulation, Kerr-lensing and diffraction. 
This simulation considers $2+1$ dimensions. 
The effect of the ultrasound pressure field is incorporated by adding a spatially dependent phase every propagation step $\d z$, leading to an expression of the optical electric field at $z+\d z$ of 
\begin{equation}
    E \left(\omega, y, z + \d z\right) = E\left(\omega, y, z\right) \times \exp \left\lbrace \iu k \Delta n\left(\omega,y,z\right) \d z \right\rbrace,
\end{equation}
where $E$ is the electric field, $z$ and $y$ are the spatial coordinates along the optical propagation and diffraction axes, respectively, $\omega$ the optical frequency, $k$ the wave vector and $\Delta n \left(\omega, y, z\right)$ is the refractive index modulation amplitude. 
In our simulation, we propagate the laser pulse seven times through the simulated sound field, reflecting the experiment. 
The parameters used in the simulation are listed in Extended Data Table \ref{ed:simparams}. 

\subsection{Detection Methods}\label{methods:optical}
\subsubsection{Optical}

We employ several detection methods for beam quality and diffraction efficiency measurements. 
They rely on two different approaches for separating the diffracted and transmitted beams: 
For efficiency and beam quality ($M^2$) measurements, we separated the diffracted order from the transmitted order using a D-shaped mirror placed approximately in the focal plane of a lens. 
Additionally, the weak higher diffraction order ($m = +2$) is removed using a knife edge. 
Although the energy content of the $m = +2$ beam is very low, it would still distort the beam quality measurement. 
The beam quality parameter $M^2$ was determined using a commercial beam profiler (Ophir-Spiricon M2-200s).
The efficiency is measured by comparing the power of diffracted and transmitted orders using photodiodes after separation of the orders. 
This approach is verified by placing a camera in the geometrical focus of a lense and comparing the integrated pixel brightness of the $m = +1$ diffracted order to all orders. 
Finally, a lens telescope collimates and magnifies the output to a $\sfrac{1}{\mathrm{e}^2}$-radius of \SI{3.3}{\milli\meter}.
The beam profile images as displayed in Fig. \ref{fig1}\textbf{d} are measured using a fibre-coupled acousto-optic modulator as a fast shutter spliced into the delivery fibre before entering the gas-AOM setup to mitigate temporal averaging over multiple modulation cycles of the acoustic field. 
The fibre AOM gate time is set to \SI{72}{\nano\second} and thus much shorter than the ultrasound field period (about \SI{2}{\micro\second}) and the camera exposure time. 
For the low optical power tests, an NKT ORIGAMI 0-10LP custom model fibre-coupled laser source is used.

The time delay between the camera exposure window and the acoustic wave can be adjusted with a delay generator (Quantum Composers Model 9514). 
For Fig. \ref{fig1}\textbf{c--d} and for the $M^2$ measurements, the delay is adjusted to maximum diffraction efficiency, resembling measurements at the peak of the ultrasound field cycle. 
By adjusting the delay for the $M^2$ measurement, we are able to show that the beam quality is independent of the exact delay setting, and thus independent of the ultrasound amplitude. 
An in-depth explanation of the temporal measurement characteristics is provided in the Methods \ref{methods:temporal} below. 
For the high optical power experiment, an AMPHOS5000 amplifier was employed.
Here, no fast shutter supporting the peak power was available. 
Thus, the beam profile is measured using a camera after \SI{16.5}{\meter} of propagation with the beam path folded over multiple plane mirrors, leading to spatial separation of transmitted and diffracted orders. 
Before this propagation and immediately after the AO-interaction, a weak portion of the beam is sampled. 
After the propagation, the diffracted and transmitted orders separate and are selected using a beam aperture. 
The beam profiles are measured using a camera and a commercially available telescope (Thorlabs BE05-UVB). 
The camera, telescope, and diaphragm settings and distances are identical for all three beam profiles (Fig. \ref{fig3}\textbf{a--e}), only the final guiding mirror was tilted to select a single diffraction order. 

\subsubsection{Acoustic}\label{sec:acoustic}
For the measurements of the travelling wave configuration, an optical microphone was used: XARION Eta100 Ultra membrane-free optical microphone \supercite{fischer_optical_2016}, specified up to \SI{180}{\deci\bel}.
It was placed \SI{1}{\centi\meter} over the UT with the sensor orientated at an angle of \SI{45}{\degree} to avoid formation of a standing wave.

\subsection{Temporal Characteristics}\label{methods:temporal}

\begin{figure}[H]
    \begin{center}
        \includegraphics{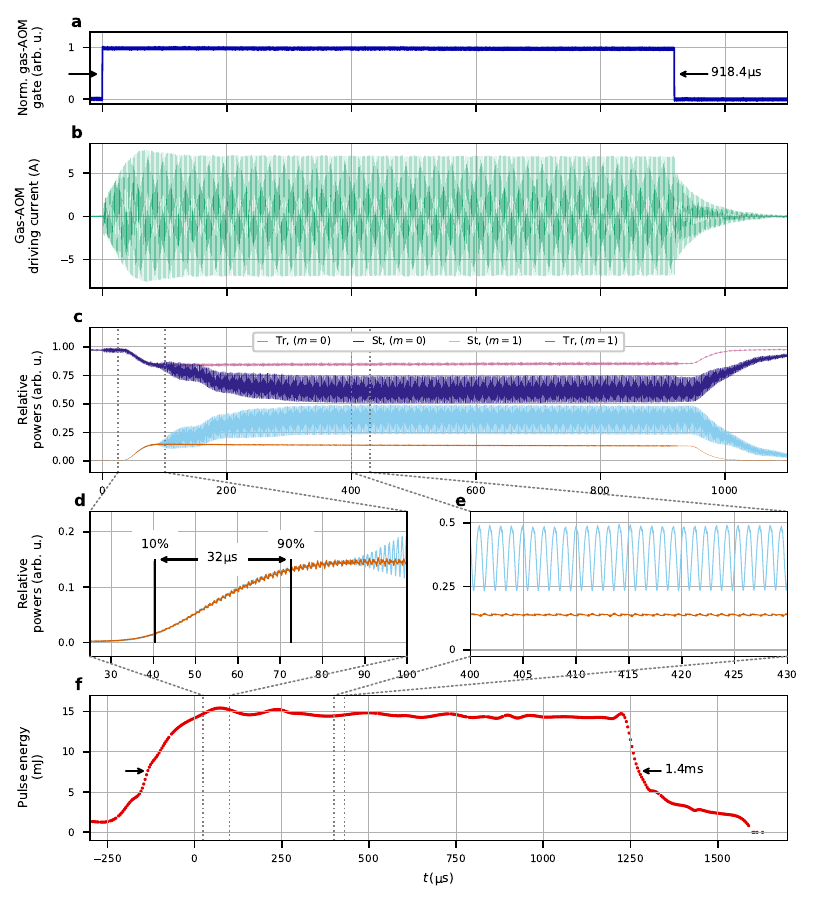}
        \caption{\label{fig:ed3}\textbf{Temporal characteristics of key parameters used for the reported experiments.} 
        \textbf{a}, Gate for the gas-AOM driver. 
        \textbf{b}, Gas-AOM driver current. 
        \textbf{c--e}, Relative powers of diffracted ($m=+1$) and transmitted ($m=0$) orders of both standing (St., light/dark blue lines) and travelling (Tr., orange/purple lines) wave configurations. 
        \textbf{f}, The reconstructed optical burst shape of the gigawatt-scale laser system. 
        The red dots indicate individual pulses used for the pulse energy calculations.}
    \end{center}
\end{figure}

The main signals relevant for our experiment display rather complex temporal gate characteristics. 
The UT operates in bursts in order to mitigate thermal effects and potential damage.
The burst frequency is \SI{5}{\hertz}, the burst duration  about \SI{1}{\milli\second}. 
Furthermore, the low power experiment is optically gated, and the high-power laser operates in bursts as well. 
Extended Data Fig. \ref{fig:ed3} details the temporal characteristics over these bursts for both experiments. 
Extended Data Fig. \ref{fig:ed3}\textbf{a} displays the gate of the gas-based AOM applied to the AOM driver, set to a duration of \SI{918}{\micro\second},corresponding to \num{450} acoustic cycles at \SI{490}{\kilo\hertz}.
Extended Data Fig. \ref{fig:ed3}\textbf{b} shows the driver current, reaching \SI{\pm 7.5}{\ampere} at a supply voltage of \SI{16}{\volt}. 
This data is measured at reduced supply voltage to avoid saturation of the 
current probe. 
Extended Data Fig. \ref{fig:ed3}\textbf{c} displays the time evolution of the relative powers contained in the $m = 0$ and $m=+1$ orders for both configurations with and without acoustic reflector.
The build-up of a standing wave after the first pass of the acoustic wave is visible when the reflector is used. 
At approximately the fifth acoustic pass, an equilibrium is reached. 
The signal is slightly deteriorating in time due to discharge of the capacitor banks of the AOM driver as seen in the behaviour of the driver current. 
Extended Data Figs. \ref{fig:ed3}\textbf{d} and \textbf{e} show expanded views of Extended Data Fig. \ref{fig:ed3}\textbf{c} containing the initial build-up phase used to extract the 10\%-90\% rise time (\SI{30}{\micro\second}) as well as the steady-state signal later in the burst. 
The beam diameter of $2 w = \SI{6.6}{\milli\meter}$ gives a lower bound for the rise time, which for air ($V=\SI[per-mode=symbol]{343}{\meter\per\second}$) is \SI{12}{\micro\second}, assuming an instantaneous transducer response. 
The rise time is thus limited by the transducer rise time.
The data agrees with the transducer rise time extracted from the temporal build-up of the driver current (Extended Data Fig. \ref{fig:ed3}\textbf{b}). 
Finally, Extended Data Fig. \ref{fig:ed3}\textbf{f} displays the reconstructed burst shape of the high-power laser, displaying both the peak pulse energy and the relatively constant burst. 
Through the delay generator, we synchronized the gas-AOM gate to the laser burst.

\subsection{Parameter Regime Estimation}\label{methods:parameter}

This section explains in detail how the parameter regimes of Fig. \ref{fig3}\textbf{f} are estimated. 
Here, the spectral and peak power regimes in which the gas-based AOM and potentially other gas-based photonic devices may be operated is visualized in the range from below \SI{0.1}{\nano\meter} (or photon energy $>\SI{10}{\kilo\electronvolt}$) up to \SI{15}{\micro\meter} and for optical peak power ranges from about \SI{100}{\watt} up to \SI{10}{\tera\watt}. 
We display the operation regimes for the gas-phase media ambient air (blue), He (yellow) and the wide-spread crystalline AO media quartz (purple) and TeO$_2$ (green), indicated using coloured areas, fading out towards unusable regimes. 
The displayed spectral limits are defined considering linear absorption in the medium using reported reference data:
Ref. \supercite{naval_air_warfare_center_atmospheric_2013} for ambient air, Ref.\supercite{henke_x-ray_1993} for He in the extreme ultraviolet and Ref. \supercite{sansonetti_handbook_2005} elsewhere, Ref. \supercite{saito_absorption_2000,linstrom_nist_1997} for quartz and Ref. \supercite{kim_linear_1993} for TeO$_2$. 
In wavelength ranges where significant absorption occurs, the colour is transparent. 
In the case of TeO$_2$, due to the lack of transmission data for wavelengths longer than \SI{3.5}{\micro\meter}, this area is marked with a dotted line. 

We estimate a limit for the optical peak power using the critical power, a lower bound for the power at which self-focusing and beam collapse will occur after sufficiently long propagation. 
The critical power for a Gaussian beam profile can be calculated as \supercite{fibich_critical_2000}:
\begin{equation}
    P_\mathrm{critical} \approx \frac{0.151 \lambda^2}{n\left(\lambda\right) n_2}.
\end{equation}
The critical power depends on the optical wavelength $\lambda$, the wavelength-dependent refractive index $n\left(\lambda\right)$ and the nonlinear refractive index $n_2$. 
This value is independent on beam size, so the applicable range may be extended beyond the critical power by limiting the propagation length in the medium and increasing beam size. 
For this reason, we choose the critical power as a reference for the limit, but fade the colour out in the range $\left[P_\mathrm{critical}; 10 P_\mathrm{critical}\right]$. 

The used wavelength-dependent refractive indices are taken from Ref. \supercite{peck_dispersion_1972} for air, Refs. \supercite{huber_refractive_1974,ermolov_supercontinuum_2015} for He, Refs. \supercite{malitson_interspecimen_1965,tan_determination_1998} for quartz and Ref. \supercite{uchida_optical_1971} for TeO$_2$. 
For He, the refractive index for $\lambda > \SI{1}{\micro\meter}$, we assume $n$ to be constant due to the lack of reference data. 
The nonlinear indices were assumed to be constant over the displayed wavelength range, and taken from Ref. \supercite{nibbering_determination_1997} for air, Ref. \supercite{bree_method_2010} for He, Ref. \supercite{milam_review_1998} for quartz and Ref. \supercite{kim_linear_1993} for TeO$_2$. 
This assumption is not justifiable at very high photon energies where a nonlinear index was not available. 
This corresponding parameter region therefore does not display a clear peak power-bound as indicated the dotted mark.

\subsection*{Data Availability}

All data and materials used in the analysis are available in the Main text, Methods, Extended Data or from the authors upon reasonable request.

\subsection*{Code Availability}

All code used in the analysis is available from the authors upon reasonable request.

\printbibliography[title={Methods References},segment=2,heading=subbibliography]
\end{refsegment}

\section{Extended Data}\label{sec:extended_data}

\begin{table}[ht!]
    \centering
    \caption{\label{ed:simparams}Parameters used for the numerical simulations of the gas-phase AO interaction.}
    \begin{tabular}{@{}ll@{}}
        \toprule
        Simulation parameter & \\
        \midrule
        \multicolumn{1}{c}{Optical}&\\
        \cmidrule{1-1}
        $\sfrac{1}{\mathrm{e}^2}$ beam radius & \SI{3.25}{\milli\meter}\\
        Optical wavelength & \SI{1030}{\nano\meter}\\
        Pulse duration & \SI{500}{\femto\second}\\
        Pulse energy & \SI{10}{\milli\joule}\\
        Number of passes over sound field & \num{7}\\
        \multicolumn{1}{c}{Acoustic}&\\
        \cmidrule{1-1}
        AO interaction length per pass & \SI{70}{\milli\meter}\\
        Free space propagation per pass & \SI{80}{\milli\meter}\\
        Ambient temperature & \SI{21.1}{\celsius}\\
        Relative humidity & $\num{42.5}\%$\\
        Acoustic frequency & \SI{490}{\kilo\hertz}\\
        \bottomrule
    \end{tabular}
\end{table}

\end{appendices}

\end{document}